\documentclass[final,5p,times,twocolumn,number]{elsarticle}

\usepackage{amssymb}
\usepackage{lipsum}
\usepackage{longtable}
\usepackage{amsmath, amssymb}
\usepackage{graphicx}
\usepackage{comment}
\usepackage{dcolumn}
\usepackage{placeins}
\usepackage{float}
\biboptions{sort&compress}%\usepackage[authoryear]{natbib}

\usepackage{color}
\renewcommand{\BibitemShut}[1]{}

\usepackage{xcolor}

\usepackage{hyperref}
 \hypersetup{
        colorlinks=true, 
        urlcolor=blue,    % Colour for external hyperlinks
        linkcolor=blue,   % Colour of internal links (e.g., sections, figures)
        citecolor=blue    % Colour of citations
    }
\newcommand{\kms}{km\,s$^{-1}$}

\journal{Physics Letters B}

\begin{document}

\begin{frontmatter}



\title{Bayesian Learning of (n,p) Reaction Cross Sections with Quantified Uncertainties}


\author[first]{Arunabha Saha}
\ead{arunabhaphy@gmail.com}
\affiliation[first]{organization={Department of Physics, ICFAI University Tripura},%Department and Organization
            addressline={Kamalghat}, 
            city={Mohanpur},
            postcode={799210}, 
            state={Tripura},
            country={India}}

\author[second]{Songshaptak De}
\ead{songshaptak.de@ijs.si}
\affiliation[second]{organization={Jo\v{z}ef Stefan Institute},%Department and Organization
            addressline={Jamova 39}, 
            %city={Sachivalaya Marg, Sainik School},
            postcode={1000}, 
            state={Ljubljana},
            country={Slovenia}}

\begin{abstract}
Accurate neutron-induced $(n,p)$ reaction cross sections are essential for applications in nuclear energy, radionuclide production, materials studies, and nuclear astrophysics. However, experimental data remain sparse for many isotopes, and evaluated nuclear data libraries can show systematic deviations from available measurements. We develop a Bayesian neural network (BNN) residual learning model, denoted \texttt{BNN-R5}, to improve $(n,p)$ reaction cross-section predictions. The model uses five physically motivated nuclear descriptors and does not employ experimental or evaluated cross-section values as input features. Rather than predicting the cross sections directly, \texttt{BNN-R5} learns the log-space residual between the evaluated TENDL-2023 data and experimental measurements, thereby providing a data-driven correction to the evaluated library. The model is trained using stochastic variational inference, which provides predictive mean values together with Bayesian uncertainty estimates. Across a broad range of target nuclei, the corrected cross sections generally show improved agreement with experimental data and outperform the original TENDL-2023 evaluations. Feature-importance analysis using SHapley Additive exPlanations (SHAP) identifies the pairing term $\delta$ as the most influential descriptor, followed by the excitation-energy variable $\ln(\Delta E)$ and the neutron number $N$, while the proton number $Z$ has the smallest overall influence. These results demonstrate that Bayesian residual learning provides a robust and interpretable framework for improving evaluated nuclear data and predicting reaction cross sections in data-sparse regions of the nuclear chart.

\end{abstract}



\begin{keyword}
%% keywords here, in the form: keyword \sep keyword, up to a maximum of 6 keywords
Neutron-induced reactions \sep Bayesian neural networks \sep Residual learning \sep Nuclear reaction cross sections% \sep Evaluated nuclear data 



\end{keyword}

\end{frontmatter}

%\tableofcontents



\section{Introduction}
\label{introduction}

\label{sec:level1}

Nuclear reaction cross sections play a fundamental role in understanding the microscopic mechanisms governing the interaction between incident nucleons and target nuclei. Reliable knowledge of these quantities is essential for applications in nuclear astrophysics, reactor design, transmutation studies, and nuclear data evaluations. In particular, (n,p) reactions are of considerable importance because they probe charge-exchange processes that provide insights into isospin-dependent nuclear structure effects and nucleon-–nucleon interactions~\cite{Chadwick2011,Kalbach1986}. Moreover, accurate (n,p) cross sections are crucial for the production of medically important radionuclides~\cite{Jafir2025}. The cross sections of (n,p) reactions are also required for design calculations of nuclear reactors. In fusion reactors, bombardment of fast neutrons on the elements of first wall, structural and blanket components of the reactors leads to (n,p), (n,$\alpha$), (n,n'$\alpha$) and (n,n'p) reactions. These reactions produces hydrogen and helium gases at different locations in the reactor wall. Owing to this, the mechanical properties of the reactor walls may get deteriorated. This has made the development of radiation-resistant materials for the reactors extremely important in the recent years. For this purpose, the accurate measurement of cross sections for the production of hydrogen and helium induced by neutrons with energy up to 20 MeV in the reactor materials are necessary. These cross sections are also necessary to estimate displacement per atom~\cite{Kondo2006} and extent of nuclear heating.

Despite the continuous efforts in experimental nuclear physics, available cross-section data remain incomplete, particularly for isotopes far from stability or at high incident energies. Consequently, theoretical nuclear reaction models such as those based on the Hauser-–Feshbach formalism implemented in codes like TALYS-2.0~\cite{Koning2023} and the evaluated data libraries such as TALYS Evaluated Nuclear Data Library (TENDL-2023)~\cite{Koning2019} and Evaluated Nuclear Data File (ENDF/B-VIII.1)~\cite{ENDFBVIII1}~—are routinely employed to supplement missing experimental information. However, these models often depend on multiple nuclear parameters like optical model potentials, level densities, and $\gamma$-strength functions, which may introduce systematic uncertainties.

In recent years, machine learning (ML) approaches have emerged as powerful alternatives for modeling complex nonlinear relationships within nuclear data. For example, it has demonstrated tremendous success in studying heavy ion collisions~\cite{Wang2023,Li2022,Pang2018,Wei2023,Wang2021}, properties of strongly interacting quantum chromodynamics (QCD) matter~\cite{Zhou2024,Kuttan2021,Li2023,Du2022}, nuclear spallation and projectile fragmentation reactions~\cite{Peng2022,Ma2022,Song2023}, nuclear fission~\cite{Song2023,Lovell2020,Wang2019,Wang2021a,Wang2022}, nuclear masses~\cite{Mumpower2022,Lovell2022,Niu2018,Gao2021,Ming2022,Wu2021,Zhao2022,Le2023}, $\beta$ decay half-lives and energy~\cite{Peng2022a,Gao2023,Munoz2023}, $\alpha$ decay~\cite{Li2022a,Ma2023,Yuan2022,Jin2023}, charge radius of atomic nuclei~\cite{Dong2023,Su2023,Wu2020,Dong2022}, nuclear density distribution~
\cite{Yang2021,Shang2022}, prediction of heavy-ion fusion cross sections~\cite{Li2024} and evaporation residual cross sections for superheavy nuclei~\cite{Zhao2022a}.

In comparison to traditional machine learning methods, the Bayesian Neural Network (BNN) approach offers distinct advantage and is one of the most important approaches for studying nuclear physics because it automatically avoids overfitting and quantification of uncertainties in model predictions. In recent years, significant improvements have been found using BNN in the prediction of nuclear mass~\cite{Utama2016a}, uncertainty of nuclear level density~\cite{Wang2024}, quantified predictions of proton and neutron separation energies~\cite{Neufcourt2020}, nuclear charge radii~\cite{Utama2016b}, prediction of (n,p) reaction cross sections~\cite{Li2024a}, and drip line locations~\cite{Neufcourt2018,Neufcourt2019}.

The present work employs a BNN framework to predict the excitation functions of neutron-induced $(n,p)$ reaction cross sections over a broad range of target nuclei. Unlike many existing machine-learning approaches reported in the literature~\cite{Li2024a,KHALI}, which utilize previously measured or evaluated cross-section values as input features, the proposed model relies exclusively on physically motivated nuclear descriptors. Consequently, the model learns the underlying systematics of $(n,p)$ reactions directly from nuclear properties, enabling reliable predictions even for nuclei where little or no experimental cross-section information is available.

Instead of directly predicting the reaction cross sections, the BNN is trained to learn the residuals between the evaluated TENDL-2023 cross sections and the corresponding experimental data. This residual-learning strategy is motivated by the fact that modern evaluated nuclear data libraries already incorporate well-established reaction physics and reproduce the overall trends of excitation functions. By learning only the systematic deviations from TENDL-2023, the BNN focuses on correcting deficiencies arising from theoretical approximations and evaluation uncertainties, thereby reducing the complexity of the learning task and improving predictive accuracy. The corrected cross sections are obtained by adding the predicted residuals to the TENDL-2023 values, resulting in predictions that consistently outperform the original TENDL-2023 evaluated data and exhibit improved agreement with experimental measurements. The predictive capability of the proposed framework is further assessed through comparisons with the evaluated nuclear data libraries TENDL-2023 and ENDF/B-VIII.1.

To further investigate the influence of individual nuclear parameters on the predicted cross sections, the SHapley Additive exPlanations (SHAP) approach~\cite{Lundberg2017} is employed. SHAP provides a consistent and interpretable framework for quantifying the contribution of each input feature to the model predictions, thereby offering valuable physical insight into the factors governing $(n,p)$ reaction cross sections. Overall, the proposed framework establishes a physics-guided, data-driven methodology that complements conventional nuclear reaction models while significantly improving the predictive accuracy of evaluated nuclear data, thereby contributing to the development of next-generation nuclear data libraries.

The rest of the paper is structured as follows. In Sect.~\ref{sec:level2}, we have discussed the methodology adopted in the present work, including the data preparation and BNN model implementation. In Sect.~\ref{sec:level3}, the results are shown and discussed in detail. Finally, in Sect.~\ref{sec:level4}, we have concluded our work.

%%%%%%%%%%%%%%%%%%%%%%%%%%%%%%%%%%%%%%%%%%%%%%%%%%%%%%%%%%%%%%%%%%%%%%%%%%%%%%%%
\label{Sect.I}

\section{\label{sec:level2}Methodology}

   A Bayesian Neural Network extends the standard artificial neural network (ANN) framework by treating the network’s parameters, weights and biases, as random variables with associated probability distributions, rather than fixed values. Unlike a conventional ANN that provides a single best-fit output for a given input, a BNN yields a probability distribution over outputs. From this distribution, one can extract both a central value and credible intervals that serve as uncertainty bounds. This feature is particularly important in nuclear data modeling, where experimental measurements are often sparse and carry significant uncertainties. As a result, BNNs seem to be an attractive choice for modeling (n,p) reaction cross sections. A detailed exposition of the theoretical foundations and development of BNNs lies beyond the scope of this study; for a comprehensive treatment, see Refs.~\cite{Bishop1995, Neal1996, Haykin1999, Vapnik1998}. In the following subsections, we describe the dataset and the BNN model employed in our analysis.

\subsection{\label{sec:level2a}Data Preparation}

    Our dataset consists of measured (n,p) reaction cross-sections for a broad range of target nuclei ($8 \lesssim Z \lesssim 77$) and incident neutron energies up to $\sim$20~MeV. These data were drawn from evaluated nuclear data library (i.e., ENDF/B-VIII.1), ensuring a comprehensive coverage of isotopic chains and energy points. Each data sample is described by an input-–output pair $(\mathbf{x}, r)$, where the target quantity $r$ is the log-space residual defined in Eq.~\ref{eq:residual} as,
    \begin{equation}
        r = ln(\sigma_{\rm ENDF}) - ln(\sigma_{\rm TENDL})
        \label{eq:residual}
    \end{equation}
    where $\sigma_{\text{ENDF}}$ is the evaluated experimentally observed cross-section (in barns) and $\sigma_{TENDL}$ is denoting the TENDL-2023 theoretical cross section data (in barns). For incident neutron energies below 20 MeV, the cross sections of (n,p) reactions are typically less than 1 barn and can vary over several orders of magnitude. Therefore, we have used the logarithmic values of the (n,p) reaction cross sections and all target values are expressed in log-space. The reconstructed BNN cross section in log-space is then recovered using the predicted residual, $r$ as given in Eq.~\ref{eq:reconstruction},
    \begin{equation}
        ln({\sigma}_{\rm BNN}) \;=\; ln(\sigma_{\rm TENDL}) + r.
        \label{eq:reconstruction}
    \end{equation}
    The input feature vector $\mathbf{x}$ is designed to encode the nuclear properties and reaction conditions. We design the input features to incorporate both basic nuclear properties and key physical quantities identified in previous studies~\cite{Li2024a}. The input vector is $\mathbf{x} = (Z,\;N,\;\delta,\;\ln(\Delta E),\;(N-Z)/A)$. Here $Z$ and $N$ are the proton and neutron numbers of the target nucleus, $\delta \in \{+1,\,0,\,-1\}$ is the pairing correction for even--even, odd-$A$, and odd--odd nuclei, respectively, capturing odd--even staggering effects in the cross sections, $\ln(\Delta E)$ is the logarithm of the energy offset above the reaction threshold where $\Delta E = E - E_{\rm thresh}$ yields more uniform excitation-function patterns across isotopes (values below $10^{-12}$~MeV are clipped before taking the logarithm), and $(N-Z)/A$ is a dimensionless isospin asymmetry parameter. Among these features, $\delta$ and $Z$ are treated as categorical variables. $\delta$ is encoded as a discrete class index, while the proton number is discretised into four shell-motivated bins ($Z \leq 20$, $20 < Z \leq 40$, $40 < Z \leq 60$, $Z > 60$), broadly corresponding to the major proton shell-closure regions. Both categorical variables are embedded into continuous learned representations via dedicated embedding layers~\cite{guo2016entityembeddingscategoricalvariables}, as described in Sect.~\ref{sec:level2b}. This separation enables the network to handle ordered numerical variables as continuous inputs, while learning distinct representations for discrete nuclear-structure effects associated with pairing and shell-motivated $Z$ regions.

    For model development, a total of 8,110 data points were randomly partitioned into training (80\%), validation (10\%), and test (10\%) subsets using a fixed random seed. The training set was used to fit the model parameters, while the validation set was employed to monitor performance and tune hyperparameters, thereby helping to prevent overfitting. The test set was used to evaluate the model’s predictive capability for (n,p) cross-sections of nuclei beyond the training domain. All continuous input features were standardised to zero mean and unit variance using statistics computed on the training set alone, and the target residual $r$ was likewise standardised prior to training.

\subsection{\label{sec:level2b}BNN Model Implementation}

%\begin{figure*}[h!]
\begin{figure*}[t]
     %\vspace{-10cm}
    \centering
    \includegraphics[width=\textwidth, trim=0 18.0cm 0 8.5cm, clip]{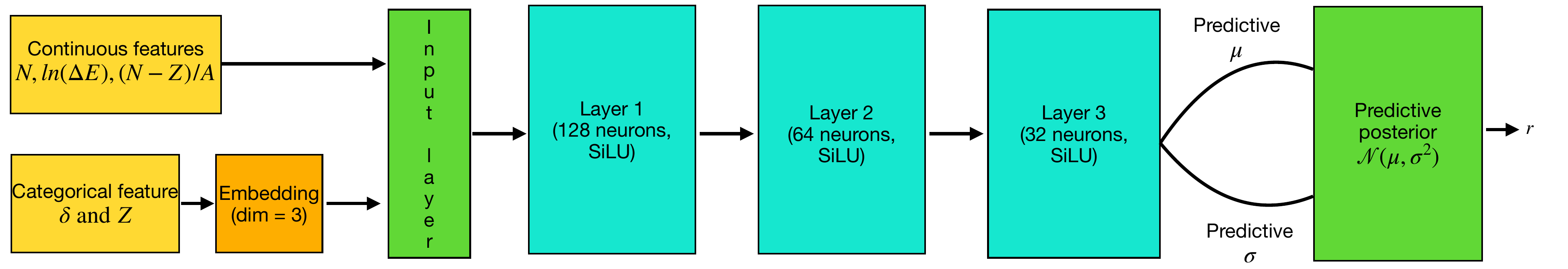}
    %\vspace{-2cm}
    \caption{(color online)~Schematic diagram of our Bayesian neural network \texttt{\texttt{BNN-R5}} used to model the residuals $r$. Continuous inputs ($N$, $\ln(\Delta E)$, $(N-Z)/A$) and an embedded categorical feature ($\delta$ and $Z$, embedding dimension 3) are concatenated in the input layer and passed through three fully connected Bayesian layers (128, 64, and 32 units, respectively, each with SiLU activation and weights treated as random variables with learned posterior distributions). The network outputs the parameters $\mu$ and $\sigma$ of a predictive Gaussian posterior $\mathcal{N}(\mu,\sigma^2)$, from which the residual $r$ is sampled.}
    \label{fig:bnn_architecture}  
\end{figure*}
\noindent

The Bayesian Neural Network (BNN) model, illustrated in Fig.~\ref{fig:bnn_architecture}, is developed as a regression architecture aimed at predicting the log-space residual $r$ defined in Eq.~(\ref{eq:residual}). We refer to this model as \texttt{\texttt{BNN-R5}}, indicating residual learning with five input features. The architecture comprises the following components.

\begin{itemize}
    \item \textbf{Input layer:} The $\delta$ index and the Z-bin index are each mapped to continuous dense representations of dimension three via trainable embedding matrices. These embeddings are concatenated with the 3 continuous features $N$, $\ln(\Delta E)$ and $(N-Z)/A$ to form an augmented input vector of total dimension $3 + 3 + 3 = 9$. 

    \item \textbf{Hidden layers:} Three fully connected layers of widths 128, 64, and 32 neurons transform the augmented input, with a Sigmoid Linear Unit (SiLU) activation function~\cite{elfwing2017sigmoidweightedlinearunitsneural} applied after each layer. SiLU\footnote{Alternative activation functions, including Exponential Linear Unit (ELU)~\cite{clevert2016fastaccuratedeepnetwork}, Gaussian Error Linear Unit (GELU)~\cite{hendrycks2023gaussianerrorlinearunits}, and Mish~\cite{misra2020mishselfregularizednonmonotonic}, were also explored during model development. However, SiLU was found to provide comparatively stable training and slightly improved predictive performance, and is therefore adopted in the present work.}  was selected for its smooth, continuously differentiable profile, which facilitates stable gradient estimation during variational inference. 
    
    \item \textbf{Output:} The final hidden representation is passed to two independent linear output heads. The first head produces the predictive mean $\mu(\mathbf{x},\mathbf{w})$ of the residual. The second head produces a raw scalar passed through the softplus function to yield a strictly positive, input-dependent noise scale $\sigma_\varepsilon(\mathbf{x},\mathbf{w}) > 0$.

\end{itemize}

\noindent The observation model is therefore 
\begin{equation}
    p(y_i \mid \mathbf{x}_i, \mathbf{w}) \;=\;
    \mathcal{N}\!\left(\mu(\mathbf{x}_i,\mathbf{w}),\;
    \sigma_\varepsilon(\mathbf{x}_i,\mathbf{w})^2\right),
    \label{eq:likelihood}
\end{equation}
where $\mathbf{w}$ collects all network parameters. Following the Bayesian framework of Bayes' theorem~\cite{bayes, Fienberg2006WhenDB, Efron2013, Berrar2019BayesTA}, the weights are assigned prior distributions, 
\begin{equation}
    P(\mathbf{w}) \;=\; \prod_{ij} \mathcal{N}(w_{ij} \mid 0,\,\tau^2),
    \label{eq:prior}
\end{equation}
with $\tau = 0.8$ for all hidden-layer and mean-head parameters. The noise-head parameters are assigned tighter priors ($\tau = 0.5$) with a bias prior mean of $-1$, encoding a weak preference for moderate initial noise estimates. The posterior distribution over weights given the training data $\mathcal{D}$ follows from Bayes' theorem:
\begin{equation}
    P(\mathbf{w} \mid \mathcal{D}) \;=\;
    \frac{P(\mathcal{D} \mid \mathbf{w})\cdot P(\mathbf{w})}
         {P(\mathcal{D})}.
    \label{eq:bayes}
\end{equation}

Since direct computation of the posterior is intractable for neural networks, we employ Stochastic Variational Inference (SVI)~\cite{JMLR:v14:hoffman13a}, approximating the true posterior by a tractable variational distribution $q(\mathbf{w}\mid\boldsymbol{\phi})$. Specifically, we adopt an AutoLowRankMultivariateNormal~\cite{ong2017gaussianvariationalapproximationfactor} guide with rank 30, which parameterises the posterior as a multivariate Gaussian with a low-rank-plus-diagonal covariance structure, capturing
inter-parameter correlations that a fully diagonal mean-field approximation would miss. The variational parameters $\boldsymbol{\phi}$ are optimised by maximising the Evidence Lower Bound (ELBO)~\cite{Blei03042017, kingma2022}:
\begin{equation}
    \mathcal{L}_{\rm ELBO} \;=\;
    \mathbb{E}_{q_\phi}\!\left[\ln p(\mathcal{D} \mid \mathbf{w})\right]
    \;-\;
    \mathrm{KL}\!\left(q(\mathbf{w}\mid\boldsymbol{\phi})\,\|\,
    P(\mathbf{w})\right).
    \label{eqn:elbo}
\end{equation}
The first term is the expected log-likelihood, measuring how well the model explains the observed data under the variational posterior. The second term is the Kullback--Leibler (KL) divergence~\cite{Kullback:1951zyt}, which regularises the model by penalising deviations of the variational distribution from the prior, thereby preventing overfitting and providing principled uncertainty estimates.

ELBO gradients are estimated using four Monte Carlo particles per optimisation step, and training proceeds with the ClippedAdam optimiser~\cite{ahn2024adammodelexponentialmoving} at a learning rate of $10^{-3}$ with gradient clipping at $\ell_2$ norm 5.0, for a maximum of 5000 epochs. An early-stopping criterion with patience of 700 epochs monitors the root-mean-square error on the validation set, evaluated every 100 epochs from 100 posterior predictive samples, with the best parameter
state restored at termination.

Once the variational posterior has been learned, predictions for unseen inputs are obtained by marginalising over $q(\mathbf{w}\mid\boldsymbol{\phi})$:
\begin{equation}
\begin{split}
p(y^* \mid \mathbf{x}^*, \mathcal{D})
&= \int p(y^* \mid \mathbf{x}^*, \mathbf{w})\,
p(\mathbf{w}\mid\mathcal{D})\,\mathrm{d}\mathbf{w} \\
&\approx \frac{1}{S}\sum_{s=1}^{S}
\mathcal{N}\!\left(\mu^{(s)},\,\sigma_\varepsilon^{(s)\,2}\right),
\end{split}
\label{eq:predictive}
\end{equation}
where $S = 1000$ weight samples are drawn at test time. The predictive uncertainty decomposes into two physically distinct contributions~\cite{H_llermeier_2021}: (i)~\textit{aleatoric uncertainty}, captured by the heteroscedastic noise
$\sigma_\varepsilon(\mathbf{x},\mathbf{w})$ and representing intrinsic variability in the evaluated data and (ii)~\textit{epistemic uncertainty}, arising from the spread of predictions across different weight samples and reflecting the model's imperfect knowledge of the underlying reaction mechanism..

To quantitatively assess model performance and facilitate comparison with TENDL-2023~\cite{Koning2019}, we define the root mean square (r.m.s.) error
\begin{equation}
    \sigma_{\rm rms} \;=\;
    \sqrt{\frac{1}{N} \sum_{i=1}^{N}
    \left( X_i^{\rm pred} - X_i^{\rm actual} \right)^2},
    \label{eq:rms}
\end{equation}
where $X_i^{\rm pred}$ denotes either the \texttt{BNN-R5} predicted $\ln(\sigma_{\rm BNN})$ or the TENDL-2023 value $\ln(\sigma_{\rm TENDL})$, and $X_i^{\rm actual} = \ln(\sigma_{\rm ENDF})$ is the corresponding ENDF/B-VIII.1 reference value.

\section{\label{sec:level3}Results and Discussions}

In this section, we present the performance of our \texttt{BNN-R5} model in predicting the experimental (n,p) cross-sections for a wide range of Z values of parent nuclei.

\begin{figure*}[th!]
   \begin{center}
      \includegraphics[scale=0.5,angle=0]{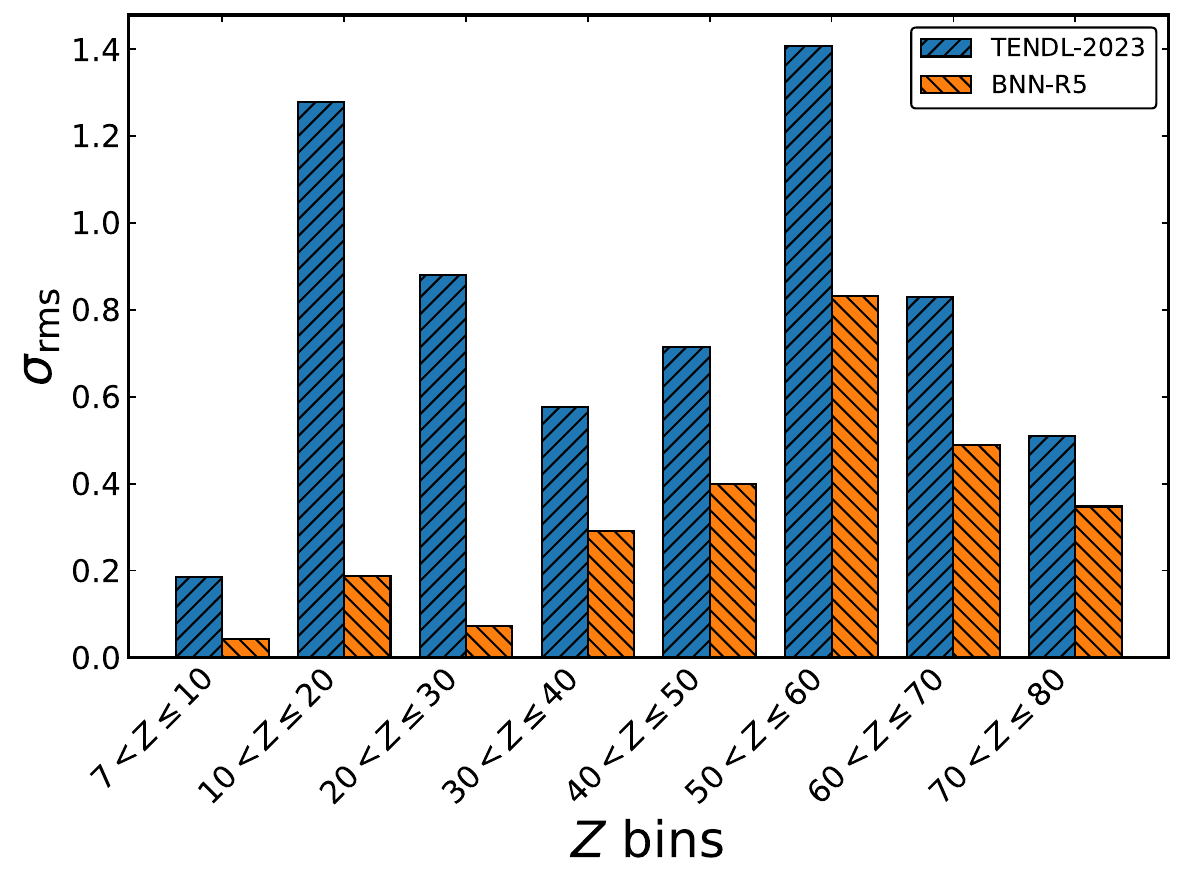}
%      \vspace{-20mm}
  \caption{(color online)Root-mean-square error ($\sigma_{\mathrm{rms}}$) of (n,p) reaction cross sections predicted by the \texttt{BNN-R5} model and those from the TENDL-2023 evaluation with respect to the available evaluated experimental data, shown as a function of the parent atomic number ($Z$) bins. The \texttt{BNN-R5} predictions generally yield lower $\sigma_{\mathrm{rms}}$ values than TENDL-2023 across all $Z$ bins, with more pronounced reductions in the intermediate and heavy mass regions, suggesting improved agreement with the evaluated experimental data.}
                \label{fig.2}
   \end{center}
\end{figure*}

Fig.~\ref{fig.2} presents the r.m.s. error metric ($\sigma_{\mathrm{rms}}$) of the (n,p) reaction cross sections obtained from the \texttt{BNN-R5} model and the TENDL-2023 evaluation relative to the evaluated experimental data, grouped according to the proton number (Z). Overall, the BNN-based approach demonstrates strong generalization capability across the nuclear chart and effectively captures systematic reaction trends without overfitting to specific mass regions.

\begin{figure*}[th!]
   \begin{center}
      \includegraphics[scale=0.4,angle=0]{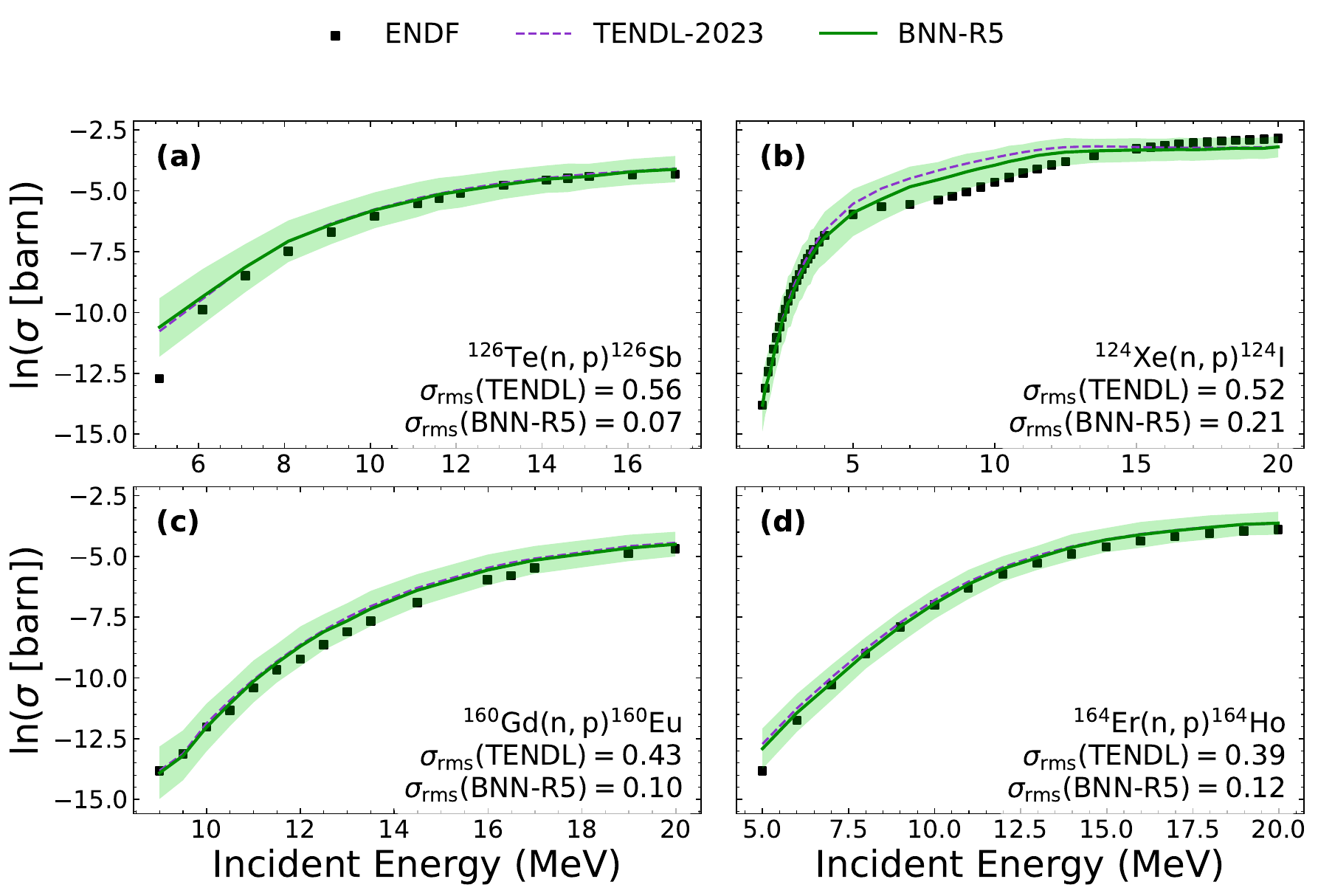}
%      \vspace{-20mm}
\caption{(color online) Comparison of the evaluated experimental data of (n,p) reaction cross sections (solid symbols) with predictions from the \texttt{BNN-R5} model (solid green lines) and the TENDL-2023 evaluation (dashed violet lines) for representative medium- and heavy-mass target nuclei: (a) $^{126}$Te(n,p)$^{126}$Sb, (b) $^{124}$Xe(n,p)$^{124}$I, (c) $^{160}$Gd(n,p)$^{160}$Eu, and (d) $^{164}$Er(n,p)$^{164}$Ho. The shaded region denotes the uncertainty band of the \texttt{BNN-R5} predicted distributions. The corresponding $\sigma_{\mathrm{rms}}$ error metric values are indicated for both TENDL-2023 and \texttt{BNN-R5} in each panel. The \texttt{BNN-R5} predictions generally follow the experimental trends over the measured energy range and yield lower $\sigma_{\mathrm{rms}}$ deviations than TENDL-2023 for these representative cases. The data are taken from the testing data set.}
                                         \label{fig.3}
   \end{center}
\end{figure*}
\begin{figure*}
   \begin{center}
	\includegraphics[scale=0.4,angle=0]{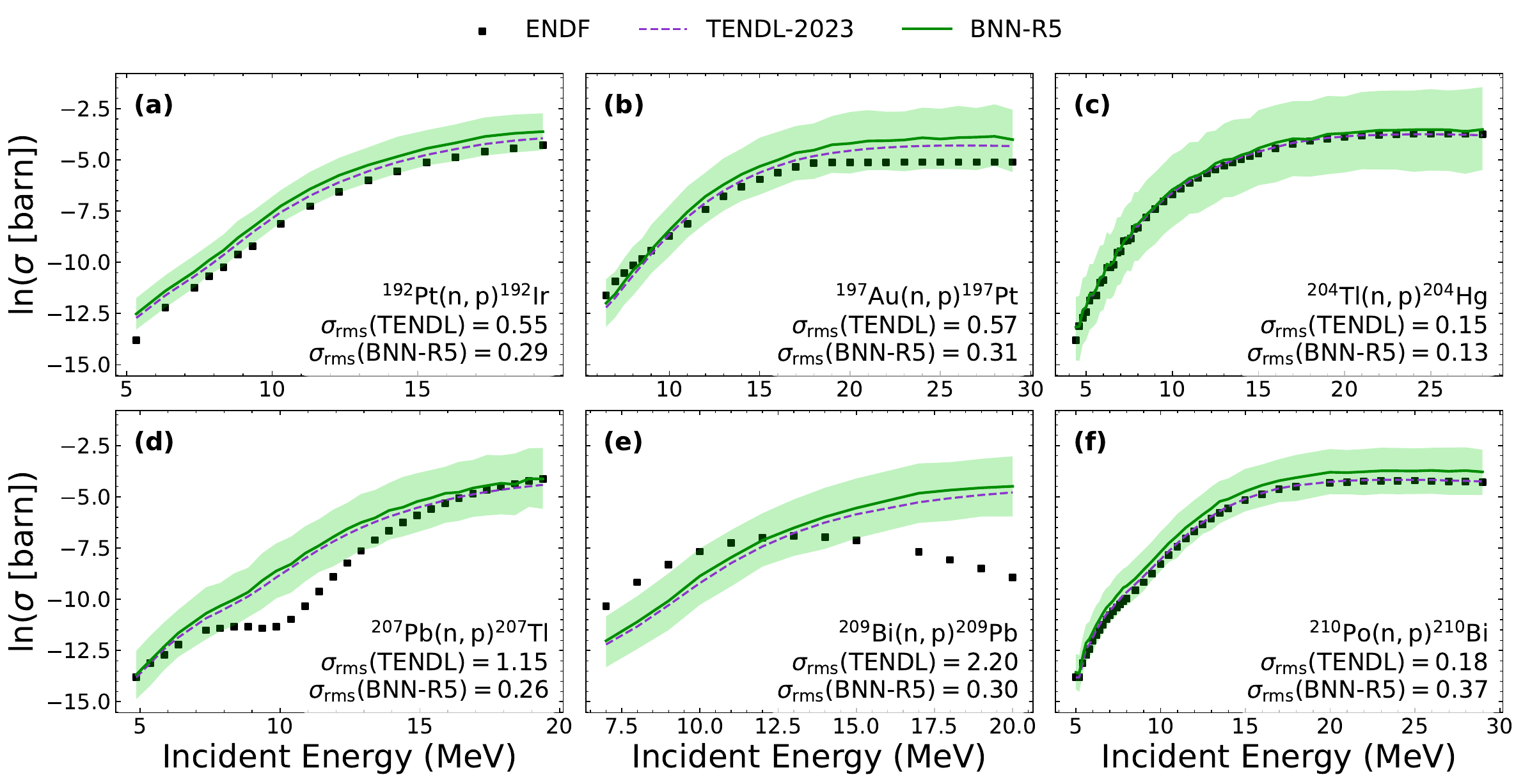}
%      \vspace{-20mm}
\caption{(color online) Same as Fig.~3, but for representative heavy target nuclei: (a) $^{192}$Pt(n,p)$^{192}$Ir, (b) $^{197}$Au(n,p)$^{197}$Pt, (c) $^{204}$Tl(n,p)$^{204}$Hg, (d) $^{207}$Pb(n,p)$^{207}$Tl, (e) $^{209}$Bi(n,p)$^{209}$Pb, and (f) $^{210}$Po(n,p)$^{210}$Bi. The \texttt{BNN-R5} model provides an accurate reproduction of the experimental excitation functions with narrow uncertainty bands. For nuclei in the Pb--Bi region, where reaction thresholds are high and experimental data are sparse, the \texttt{BNN-R5} model maintains reasonable agreement with the available measurements. The corresponding $\sigma_{\mathrm{rms}}$ values for \texttt{BNN-R5} and TENDL-2023 are also shown in each panel for comparison. The data are taken from the testing data set that has been specially prepared for heavy target nuclei. These nuclei were not included in the dataset that has been split to training, validation and testing set.}
                                     \label{fig.4}
   \end{center}
\end{figure*}

The subsequent discussion provides a detailed evaluation of the Bayesian neural network predictions against both the evaluated nuclear data and experimental measurements. The predictive performance of the \texttt{BNN-R5} model for neutron-induced (n,p) reactions has been systematically evaluated against the data from TENDL-2023 library and evaluated experimental data from the ENDF/B-VIII.1 database. The comparison is shown in Figs.~\ref{fig.3} and ~\ref{fig.4} for representative medium- and heavy-mass nuclei, respectively. The shaded green band denote the uncertainty of the BNN predictions and the green curve denote the central line.

For the medium-mass nuclei (Fig.~\ref{fig.3}), namely $^{126}$Te,
$^{124}$Xe, $^{160}$Gd, and $^{164}$Er, the \texttt{BNN-R5} model
reproduces the evaluated ENDF excitation functions with consistently good
accuracy over the entire incident-energy range. In all four cases, the
\texttt{BNN-R5} predictions closely follow the evaluated data, while the
corresponding uncertainty bands remain narrow and physically reasonable.
Compared with the TENDL-2023 evaluation, the \texttt{BNN-R5} model yields
systematically lower r.m.s.\ deviations, reducing $\sigma_{\rm rms}$ from
0.56 to 0.07 for $^{126}$Te, from 0.52 to 0.21 for $^{124}$Xe, from
0.43 to 0.10 for $^{160}$Gd, and from 0.39 to 0.12 for $^{164}$Er. The
most pronounced improvement is observed for $^{126}$Te, where the
\texttt{BNN-R5} prediction provides an excellent description of the evaluated
excitation function across the full energy range. These results
demonstrate that the Bayesian neural network effectively captures the
underlying systematics of the $(n,p)$ reaction cross sections in the
medium-mass region while providing reliable uncertainty estimates.

Extending the analysis to heavy target nuclei further demonstrates the applicability of the \texttt{BNN-R5} framework across a broader mass region. As shown in Fig.~\ref{fig.4}, the excitation functions for the (n,p) reactions on $^{192}$Pt, $^{197}$Au, $^{204}$Tl, $^{207}$Pb, $^{209}$Bi, and $^{210}$Po are reproduced with generally good agreement with the available evaluated data. The \texttt{BNN-R5} model provides a clear improvement over the TENDL-2023 evaluation for $^{192}$Pt, $^{197}$Au, $^{204}$Tl, $^{207}$Pb, and $^{209}$Bi, reducing the corresponding r.m.s. deviations from 0.55 to 0.29, 0.57 to 0.31, 0.15 to 0.13, 1.15 to 0.26, and 2.20 to 0.30, respectively. Among these, the best agreement is achieved for $^{204}$Tl, where the r.m.s. deviation is as low as 0.13. In the Pb--Bi mass region, where reaction thresholds are relatively high and experimental information is sparse, the \texttt{BNN-R5} predictions preserve the overall energy dependence of the excitation functions while maintaining physically reasonable uncertainty bands. For $^{210}$Po, however, the \texttt{BNN-R5} prediction yields a slightly larger r.m.s. deviation (0.37) than TENDL-2023 (0.18), although it continues to reproduce the overall trend of the evaluated excitation function within the estimated uncertainty band.

\begin{figure*}[th!]
   \begin{center}
      \includegraphics[scale=0.5,angle=0]{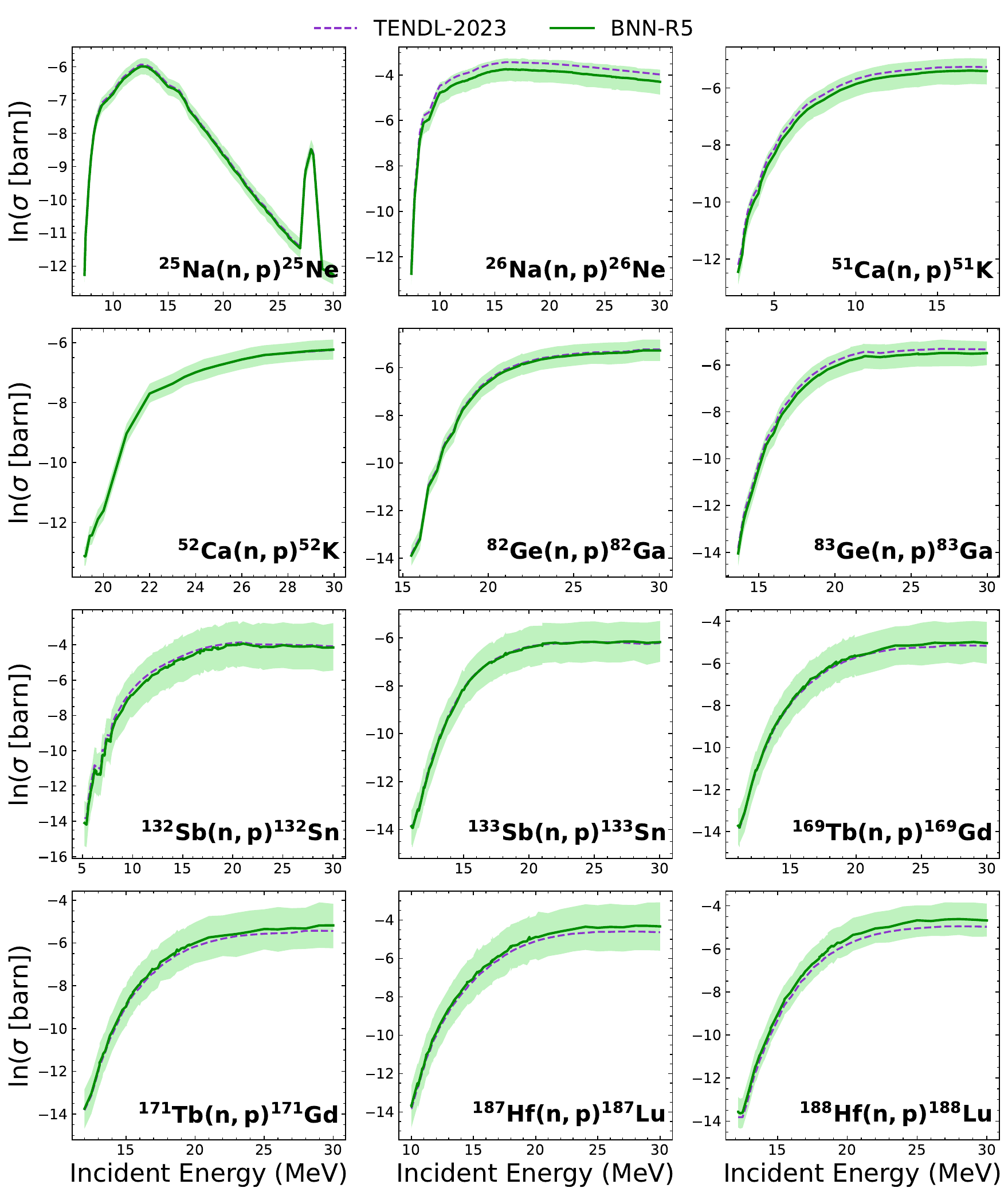}
%      \vspace{-20mm}
\caption{(color online) Comparison of the neutron-induced (n,p) reaction cross sections obtained from the \texttt{BNN-R5} model with those from the TENDL-2023 library for various target nuclei. The solid green lines represent the \texttt{BNN-R5} predicted $\ln(\sigma[\mathrm{barns}])$, while the shaded green bands denote the corresponding prediction uncertainties. The dashed violet lines indicate the TENDL-2023 evaluations. The \texttt{BNN-R5} model reproduces the overall energy dependence of the (n,p) excitation functions across most of the incident energy range. The prediction uncertainties generally increase at higher incident energies, reflecting the reduced density of experimental data and the increasing model uncertainty beyond the most constrained energy domain. The data are taken from the testing data set which has been prepared for a few selected light, medium-mass, and heavy target nuclei. These nuclei were not included in the dataset that has been split to training, validation and testing set.}
                                  \label{fig.5}
   \end{center}
\end{figure*}

The overall results highlight that the \texttt{BNN-R5} framework successfully captures the global trends of (n,p) reaction cross sections while providing statistically meaningful uncertainty estimates. This dual capability of accuracy and uncertainty quantification makes the Bayesian neural network a promising tool for supplementing evaluated nuclear data libraries, especially in regions where experimental information is sparse or uncertain.

Building upon the agreement observed for the medium- and heavy-mass nuclei, we further examine the global predictive capability of the \texttt{BNN-R5} model across representative nuclei spanning the entire nuclear chart. Figure~\ref{fig.5} presents a systematic comparison between the \texttt{BNN-R5} predictions and the TENDL-2023 evaluated $(n,p)$ excitation functions for selected light ($^{25}$Na, $^{26}$Na, $^{51,52}$Ca), medium-mass ($^{82,83}$Ge, $^{132,133}$Sb), and heavy ($^{169,171}$Tb, $^{171}$Gd, $^{187}$Hf, and $^{188}$Lu) target nuclei. Overall, the \texttt{BNN-R5} predictions closely reproduce the energy dependence of the TENDL-2023 excitation functions, accurately capturing the reaction thresholds, the rapid rise of the cross sections, and their subsequent saturation at higher incident energies. The associated predictive uncertainty bands remain relatively narrow over most of the energy range and exhibit a gradual increase at higher incident energies for some nuclei, reflecting the reduced availability of training data and the corresponding increase in epistemic uncertainty. Despite the diversity of the selected nuclei and the wide mass range considered, the \texttt{BNN-R5} predictions remain smooth and physically consistent, indicating that the Bayesian neural network has successfully learned the global systematics governing $(n,p)$ reaction cross sections and provides reliable extrapolative predictions across the nuclear chart.

\begin{figure*}[th!]
   \begin{center}
      \includegraphics[scale=0.6,angle=0]{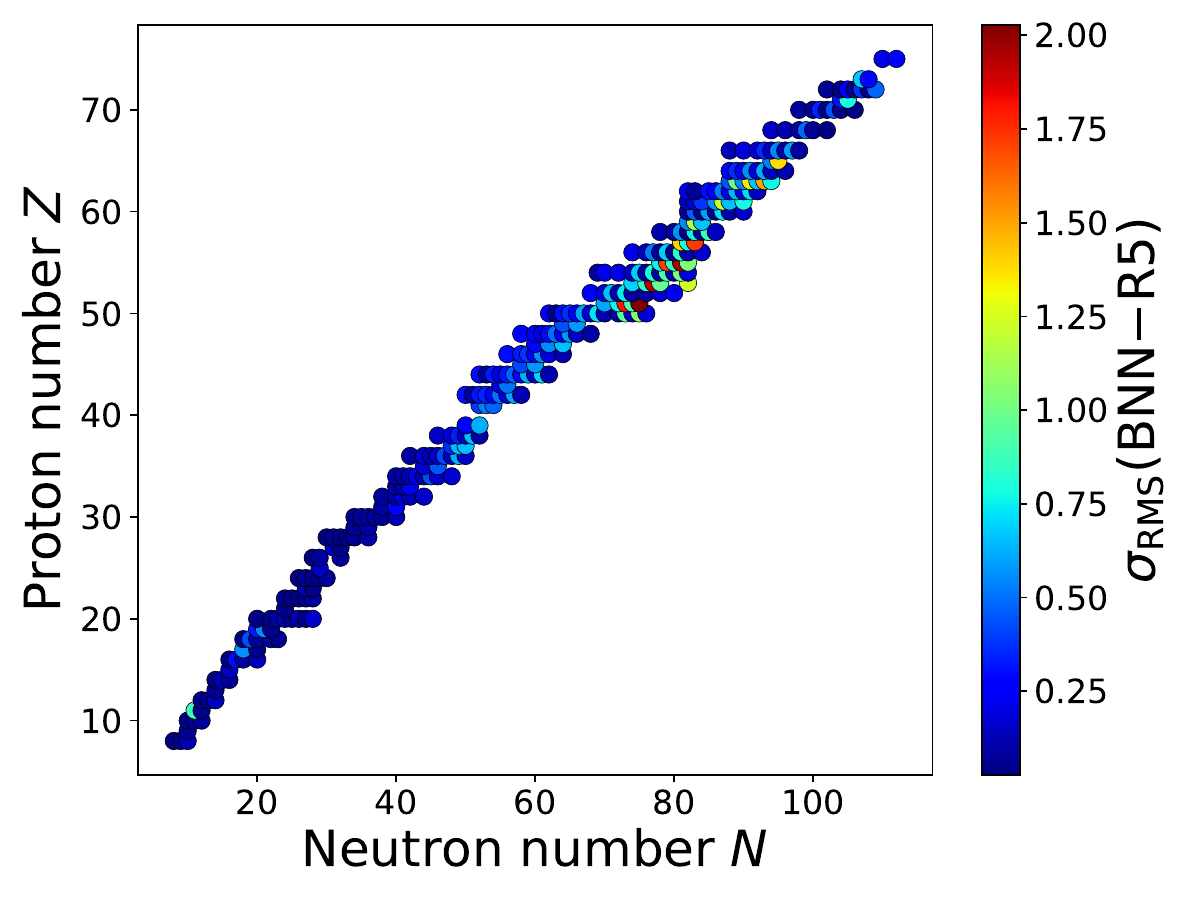}
%      \vspace{-20mm}
\caption{(color online) Distribution of the root-mean-square (r.m.s.) error ($\sigma_{\mathrm{rms}}$) between the calculated and evaluated $(n,p)$ reaction cross sections across the nuclear chart, shown as a function of neutron number ($N$) and proton number ($Z$). The color scale represents the magnitude of $\sigma_{\mathrm{rms}}$, with darker blue indicating lower prediction errors and warmer colors corresponding to larger deviations. Most nuclei exhibit low r.m.s. errors ($\sigma_{\mathrm{rms}} \lesssim 0.5$), demonstrating the overall predictive accuracy of the \texttt{BNN-R5} model. Slightly higher errors are observed for a limited number of medium- and heavy-mass nuclei, while the concentration of points along the line of stability indicates that the model effectively captures the systematic dependence of $(n,p)$ reaction cross sections on both neutron and proton numbers.}
                \label{fig.8}
   \end{center}
\end{figure*}

To provide a quantitative assessment of the predictive performance over the
entire nuclear chart, the root-mean-square (r.m.s.) deviations between the
\texttt{BNN-R5} predictions and the corresponding evaluated $(n,p)$ reaction cross
sections were calculated and are summarized in Fig.~\ref{fig.8}. The
distribution of $\sigma_{\rm rms}$ in the $(N,Z)$ plane reveals that the
majority of nuclei exhibit relatively small prediction errors
($\sigma_{\rm rms}\lesssim0.5$), indicating a consistently good agreement
between the \texttt{BNN-R5} predictions and the evaluated data. A limited number of
nuclei, primarily in the medium- and heavy-mass regions, display moderately
larger r.m.s. deviations, although these remain confined to isolated cases
rather than forming any systematic trend across the nuclear chart. The
predominantly low-error distribution demonstrates that the \texttt{BNN-R5} model
successfully captures the global dependence of $(n,p)$ reaction cross
sections on neutron and proton numbers while maintaining stable predictive
performance over a broad range of target nuclei.

\begin{figure*}[th!]
   \begin{center}
      \includegraphics[scale=0.4,angle=0]{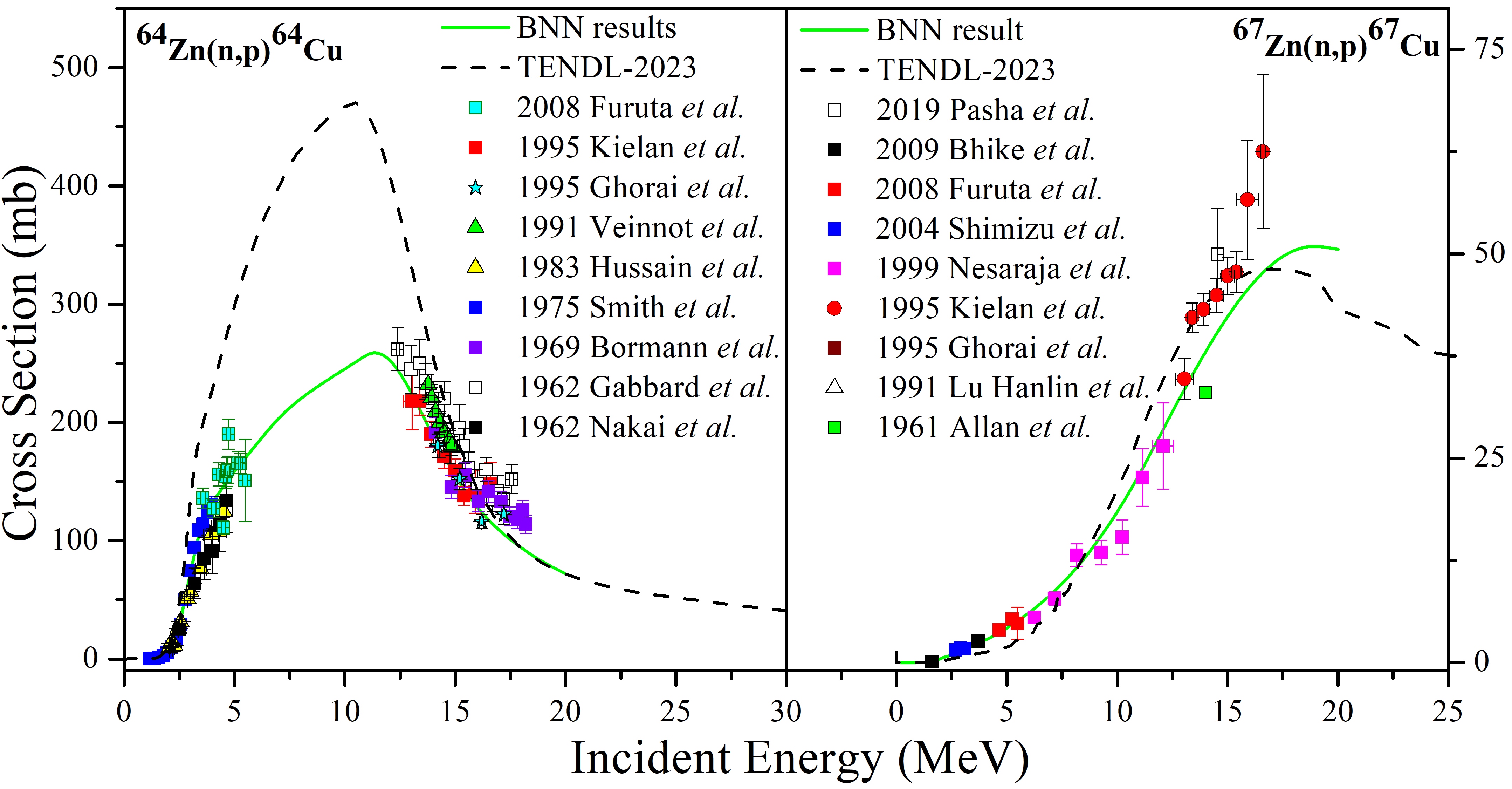}
%      \vspace{-20mm}
      \caption{(color online) Comparison of evaluated and predicted cross sections for the $^{64}$Zn(n,p)$^{64}$Cu (left) and $^{67}$Zn(n,p)$^{67}$Cu (right) reactions. The \texttt{BNN-R5} results are benchmarked against TENDL-2023 evaluations and experimental datasets from the EXFOR database, highlighting the model’s performance across the incident neutron energy range.}
                                                                 \label{fig.6}
                                                                 
   \end{center}
\end{figure*}

To further assess the predictive capability of the model, we now examine its performance for specific reactions which were not included in the training dataset. Using $^{64}$Zn and $^{67}$Zn as representative cases, Fig.~\ref{fig.6} presents a comparison between the (n,p) reaction cross sections predicted by the \texttt{BNN-R5} model and the corresponding experimental data obtained from the EXFOR library~\cite{EXFOR}. Since these reactions were not included in the training dataset, they provide an independent benchmark for evaluating the model’s predictive performance. For the sake of clarity in comparison with the experimental data, only the mean predicted values are shown in the figure. This analysis is intended to demonstrate the capability of the trained model to reliably predict cross sections for previously unseen reactions.

\begin{figure*}[th!]
   \begin{center}
      \includegraphics[scale=0.8,angle=0]{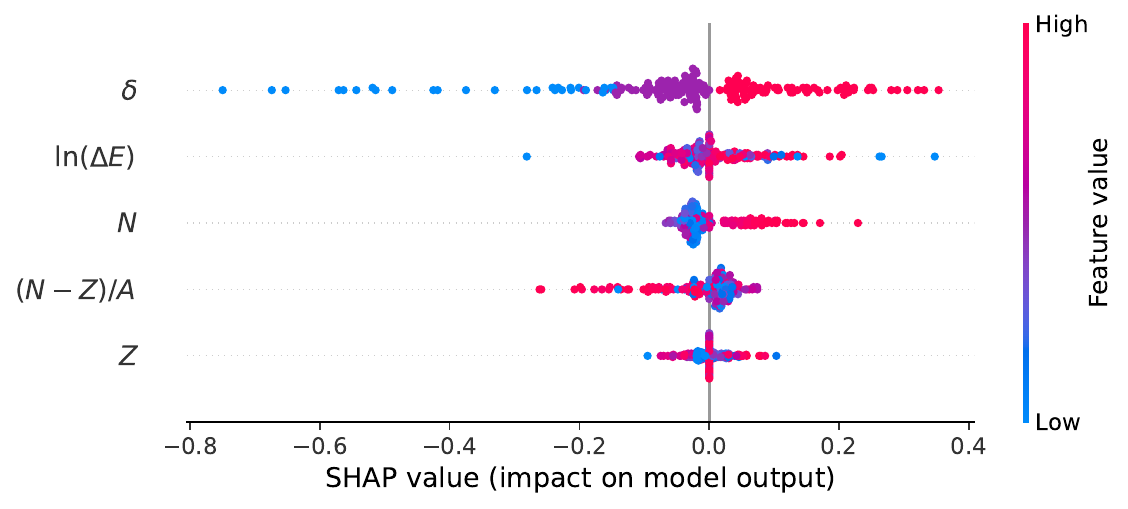}
%      \vspace{-20mm}
      \caption{(color online) Importance ranking for the input features obtained with the \texttt{SHAP} python package. Each row represents a feature, and the x axis is the SHAP value, which shows the importance of a feature for a particular prediction. Each point represents a reaction, and the color represents the feature value (with red being high and blue being low).}
                              \label{fig.7}
   \end{center}
\end{figure*}

Apart from evaluating the predictive performance of the \texttt{BNN-R5} model, it is also important to understand how the model arrives at its predictions. To interpret the contribution of different input variables, we employed the SHapley Additive exPlanations (SHAP) method, which is one of the most widely used feature attribution techniques. Figure~\ref{fig.7} presents the SHAP summary plot, illustrating the influence of each input feature on the predicted reaction cross section.

In the SHAP summary plot, the input features are arranged along the y-axis in descending order of their overall importance, namely $\delta$, ln($\Delta E$), $N$, $(N-Z)/A$, and $Z$. The x-axis represents the SHAP value, which quantifies the contribution of each feature to the model prediction. Positive SHAP values increase the predicted cross section, whereas negative values decrease it. The color scale indicates the magnitude of the corresponding feature value, with blue representing low values and red representing high values.

Among all the input variables, the pairing parameter, $\delta$, exhibits the largest spread of SHAP values, indicating that it has the strongest influence on the model output. The distribution also shows that higher values of $\delta$ (red points) predominantly contribute to positive SHAP values, thereby increasing the predicted cross section, whereas lower values (blue points) tend to reduce the prediction. The second most influential feature is ln($\Delta E$), although its SHAP values are more narrowly distributed around zero, suggesting a comparatively weaker impact. The neutron number ($N$) also contributes noticeably, with larger neutron numbers generally shifting the prediction towards higher cross sections. In contrast, the asymmetry parameter, $(N-Z)/A$, and the proton number ($Z$) exhibit relatively small SHAP values clustered near zero, implying that their influence on the model prediction is limited.

Overall, the SHAP analysis indicates that the \texttt{BNN-R5} model relies primarily on the pairing parameter $\delta$, while ln($\Delta E$) and the neutron number provide additional, though less significant, contributions. The remaining nuclear structure parameters, $(N-Z)/A$ and $Z$, play comparatively minor roles in determining the predicted reaction cross sections.

\section{\label{sec:level4}Summary}
 
In this work, we have developed a Bayesian residual-learning framework, denoted \texttt{BNN-R5}, to improve neutron-induced $(n,p)$ reaction cross-section predictions. Instead of learning the excitation functions directly, the network is trained to predict the logarithmic residual between ENDF/B-VIII.1 reference data and the TENDL-2023 evaluation. The corrected cross section is then obtained by adding the predicted residual to the TENDL-2023 baseline in log space. In this way, the model uses the global physics already encoded in evaluated nuclear-data libraries, while learning systematic deviations from the reference data. The motivation for this work arises from the need for accurate and reliable nuclear reaction data in applications such as nuclear reactor design, nuclear waste transmutation, medical isotope production, and nuclear astrophysics, where experimental measurements remain scarce for many isotopes.

The \texttt{BNN-R5} model uses five physically motivated descriptors: the target proton number $Z$, neutron number $N$, pairing term $\delta$, $\ln(\Delta E)$, and isospin asymmetry $(N-Z)/A$. The Bayesian formulation provides both central predictions and predictive uncertainties, allowing the reliability of the corrected cross sections to be assessed together with the mean prediction. To quantitatively assess the predictive performance of the \texttt{BNN-R5} model, we compare its predictions with the ENDF/B-VIII.1 reference data using the root mean square (r.m.s.) deviation defined in Eq.~(\ref{eq:rms}). The same metric is also employed to evaluate the TENDL-2023 library, enabling a direct and consistent comparison of the predictive accuracy of the two approaches. Across a broad range of target nuclei, the residual-corrected predictions show improved agreement with the ENDF/B-VIII.1 reference data compared with the original TENDL-2023 evaluation. The improvement is observed for several representative medium- and heavy-mass nuclei, although a few isolated cases remain where the correction is less effective.

The model was further tested against experimental EXFOR data for the $^{64}\mathrm{Zn}(n,p)^{64}\mathrm{Cu}$ and $^{67}\mathrm{Zn}(n,p)^{67}\mathrm{Cu}$ reactions, which were not used in the model development. The resulting excitation functions reproduce the main experimental trends, indicating that the learned residual systematics can provide useful predictions for reactions with limited available data.

To gain physical insight into the trained model, we performed a SHAP-based feature-importance analysis. The pairing descriptor $\delta$ was found to be the most influential input, followed by $\ln(\Delta E)$ and the neutron number $N$. The isospin asymmetry $(N-Z)/A$ and the proton number $Z$ have comparatively smaller overall impact. This hierarchy suggests that the model primarily exploits pairing effects while incorporating additional information from the reaction energy and nuclear structure to achieve accurate and physically meaningful predictions.

Overall, the present study shows that Bayesian residual learning offers a compact and uncertainty-aware route for improving evaluated nuclear reaction data. The approach is particularly useful as a complementary tool for nuclear-data evaluation in regions where direct measurements are sparse, while retaining interpretability through physically motivated inputs and feature-attribution analysis.

\section{Acknowledgements}

A. Saha would like to acknowledge ICFAI University Tripura for all necessary help and support. The work of S.~De~is co-funded by the European Union’s Horizon Europe research and innovation program under the Marie Sk{\l}odowska-Curie COFUND Postdoctoral Programme grant agreement No. 101081355-SMASH and by the Republic of Slovenia and the European Union from the European Regional Development Fund. 
\label{Sect.VII}

%\FloatBarrier

%\bibliographystyle{apsrev4-2}
%\bibliography{ref}

%\bibliographystyle{elsarticle-harv}
%\bibliographystyle{unsrt}
%\bibliographystyle{ieeetr}
\bibliographystyle{elsarticle-num}
\bibliography{example}

\end{document}